\documentclass{aip-cp}
\usepackage[numbers]{natbib}
\usepackage{rotating}
\usepackage{graphicx}

\newtheorem{theorem}{Theorem}

\newtheorem{definition}[theorem]{Definition}

\newtheorem{remark}[theorem]{Remark}

\begin{document}

\title{A Comparison Between the Stability Properties in a DDE Model for
Leukemia and the Modified Fractional Counterpart}

\author[aff1,aff2]{I.R. R\u{a}dulescu \corref{cor1} }
\author[aff2]{D. C\^{a}ndea}
\eaddress{doinatilivea@gmail.com}
\author[aff1,aff3]{E. Kaslik}
\eaddress{ekaslik@gmail.com}

\affil[aff1]{Institute e-Austria, Timisoara, Romania}
\affil[aff2]{Department of Mathematics and Informatics, University Politehnica of Bucharest, Romania}
\affil[aff3]{West University of Timisoara, Romania}
\corresp[cor1]{Corresponding author: nicola{\_}rodica@yahoo.com}

\maketitle

\begin{abstract}
In this paper, a delay differential equations (DDEs) model of leukemia is
introduced and its dynamical properties are investigated in comparison with
the modified fractional-order system where the Caputo's derivative is used.
The model takes into account three types of division that a stem-like cell
can undergo and cell competition between healthy and leukemia cell populations.
The action of the immune system on the leukemic cell populations is also
considered. The stability properties of the equilibrium points are established
through numerical results and the differences between the two types of approaches
are discussed. Medical conclusions are drawn in view of the obtained numerical
simulations.
\end{abstract}

\section{INTRODUCTION}

Chronic myeloid leukemia (CML) is a hematopoietic stem cell disorder characterized
by the Philadelphia (Ph) chromosome translocation. This translocation causes the
transcription of a constitutively active oncogenic enzyme with tyrosine kinase activity:
BCR-ABL (\cite{fad}). Research in the last decades has ascertained that
BCR-ABL is the cause to the CML pathogenesis and that constitutive tyrosine
kinase activity is central to BCR-ABL's ability to alter normal
hematopoietic cells into leukemic cells (\cite{dei}).

Most of the cases are diagnosed during an initial chronic phase
characterized by increased proliferation and the accumulation of immature
myeloid cells over several years. If untreated, CML inevitably progresses to
an accelerated phase and/or blast phase. The treatment of CML has been
revolutionized by the advent of specific tyrosine kinase inhibitors (TKI).
TKI selectively inhibits BCR-ABL action and, consequently, repress the
tyrosine kinase activity. The most important TKI is Imatinib, which has a
very good outcome if the treatment is not interrupted (\cite{mar}).

It is generally assumed that the immune system plays an essential role in tumor
evolution. Researches in cancer immunology imply that both innate and adaptive
immunity are implicated in the defense systems against cancer. In CML,
leukemic cells express antigens that are
immunogenic and can be recognized by cytotoxic T cells (or CD8+ T
cells), which are able to eliminate leukemic stem cells (LSCs) (\cite%
{rie}). In \cite{che}, the authors claim that some CML patients under
imatinib-induced remission develop an anti-leukemia immune response
involving both CD4+ and CD8+ T cells. New advances in cancer immunotherapy
suggest that the immune system can be used as a tool to prevent or cure
cancer \cite{top}. Consequently, immunotherapy should be considered as a
complementary therapy, in order to increase the effectiveness of the immune
response against cancer. T-cell based therapies have been shown to boost the
body's ability to fight cancers as leukemia, lymphoma and breast cancer \cite%
{del}. Therefore a model that includes T cell dynamics may serve to enhance
the potential immunotherapy treatments of CML.

There is a vast literature on mathematical modeling the normal and
pathologic hematopoiesis (\cite{adi2010}, \cite{bern}, \cite{colf}, \cite%
{col}, \cite{Halanay2014}, \cite{hal}, \cite{rad}, \cite{jtb}). In some
papers, CML models with competition between cell lines are studied using
stochastic differential equations (\cite{cat}, \cite{roe}) while in others
ordinary differential equations (ODEs) are used \cite{mac2}, \cite{foo},
\cite{mac1}, \cite{din}. However, because of the complexity of the
hematopoietic system, ODEs models cannot always capture the rich variety of
dynamics observed in real life. An intermediate approach is the use of delay
differential equations (DDEs) models, which exhibit more complex dynamical
behaviors (see, for example \cite{adi2012}).

Some models that specifically study the immune response to CML are \cite{kim}%
, \cite{ber}, \cite{nei} and \cite{moo}. In \cite{kim}, Kim et al. analyzed
a high order DDEs model to account for the role of anti-leukemia specific
response in CML dynamics and concluded that the anti-leukemia T cell
response may help maintain remission under Imatinib therapy. In \cite{ber},
the authors analyzed a DDEs model for the dynamics of CML cells and
effectors T cells considering Imatinib therapy and immunotherapy. The focus
in the paper \cite{nei} is on analyzing a DDE model in order to elucidate
the transition of leukemia from the stable chronic phase to the unstable
accelerated and acute phases. In \cite{moo}, Moore and Li devise an ODE
model and examine which parameters of the model are most important in the
success or failure of cancer remission.

The DDEs model we present in Section 2 follows the lines of the models for
leukemia developed by Mackey and collaborators (\cite{bern}, \cite{col},
\cite{puj}) and is part of a larger study of Halanay and colleagues about
CML (\cite{bad}, \cite{MESA}, \cite{Halanay2014}, \cite{hal}, \cite{rad}, \cite{jtb}).
One of the novelties in this study is the consideration of three types of
division that a stem-like cell can undergo: asymmetric division, symmetric
self-renewal and differentiation (\cite{Tomasetti2010}). The DDEs system models
the interaction between healthy cell populations, leukemic cell population and
active T cells and it consists of a strongly nonlinear five-dimensional
system of delay differential equations.

Recently, fractional differential models have been extensively used in
different fields, such as physics, biology, chemical technology and economy (%
\cite{Podlubny}, \cite{Kilbas}). Due to the fact that fractional-order
equations are naturally related to systems with memory, which exists in most
biological systems, several papers have analyzed the qualitative properties
of fractional-order biological models (see e.g. \cite{ahm}). Based on the
reasoning above and because adding a supplementary memory to the DDEs model
for CML might be more in line with the real life situation, we then present
the fractional-order DDEs model for CML and realise a comparison between the
dynamical properties of the two models.

In the following section, the delay competition model of CML, which was
first introduced in \cite{rad} (see also \cite{MESA}), is presented and
explained. Based on this model, in Section 3 we introduce the
fractional-order DDEs system for CML. Section 4 presents numerical results
and simulations, which are then discussed and interpreted from a medical
point of view in Section 5.

\section{THE DDE MODEL FOR CML AND EQUILIBRIUM POINTS}

The mathematical model presented in this section consists of a system of
strongly nonlinear differential equations with five time lags and five state
variables. The time delays represent the duration of the cell cycle ($\tau
_{1}$ and $\tau _{3})$, the time period needed for differentiation into the
leukocyte line ($\tau _{2}$ and $\tau _{4})$ and the time period needed for
the naive T cells to finish the minimal developmental program of $n_{1}$
cell divisions ($\tau _{5}=n_{1}\tau $, where $\tau $ is duration of one T
cell division). In this model it is assumed that the dynamics of healthy and
leukemic cell populations are similar and the only notable difference is in
the value of the parameters. Consequently, the subscript of the parameters
for the healthy cells is denoted with \textquotedblleft h\textquotedblright\
and the subscript for the abnormal cells is denoted with \textquotedblleft
l\textquotedblright . Also, the third and the fourth equations are similar
with the first and the second ones. Three types of division are taken into
consideration in this model: asymmetric division ($\eta _{1}$), symmetric
differentiation ($\eta _{2}$) and self-renewal ($1-\eta _{1}-\eta _{2}$).

The state variables are five cell populations: healthy stem-like and mature (%
$x_{1}$and $x_{2})$, leukemic stem-like and mature ($x_{3}$and $x_{4})$ and
anti-leukemic T cells involving both CD4+ and CD8+ T cells ($x_{5}$). The
dynamics of these populations in CML is described by the following system%
\begin{equation}
\begin{array}{lcl}
\dot{x}_{1} &=&f_{1}(x_{1},x_{2},x_{3},x_{4},x_{1\tau _{1}},x_{2\tau
_{1}},x_{3\tau _{1}},x_{4\tau _{1}})  \label{sys} \\
\dot{x}_{2} &=&f_{2}(x_{2},x_{1\tau _{2}},x_{2\tau _{2}},x_{4\tau _{2}})
 \\ 
\dot{x}_{3} &=&f_{3}(x_{1},x_{2},x_{3},x_{4},x_{5},x_{1\tau _{3}},x_{2\tau
_{3}},x_{3\tau _{3}},x_{4\tau _{3}})  \\  
\dot{x}_{4} &=&f_{4}(x_{3},x_{4},x_{5},x_{2\tau _{4}},x_{3\tau
_{4}},x_{4\tau _{4}})  \\  
\dot{x}_{5} &=&f_{5}(x_{4},x_{5},x_{4\tau _{5}},x_{5\tau _{5}})  
\end{array}%
\end{equation}%
where{%
\begin{equation}
\begin{array}{lcl}
f_{1} & = & -\gamma _{1h}x_{1}-(\eta _{1h}+\eta
_{2h})k_{h}(x_{2}+x_{4})x_{1}-(1-\eta _{1h}-\eta _{2h})\beta
_{h}(x_{1}+x_{3})x_{1}+ \\
\noalign{\medskip} &  & +2e^{-\gamma _{1h}\tau _{1}}(1-\eta _{1h}-\eta
_{2h})\beta _{h}(x_{1\tau _{1}}+x_{3\tau _{1}})x_{1\tau _{1}}+\eta
_{1h}e^{-\gamma _{1h}\tau _{1}}k_{h}(x_{2\tau _{1}}+x_{4\tau _{1}})x_{1\tau
_{1}} \\
\noalign{\medskip}f_{2} & = & -\gamma _{2h}x_{2}+A_{h}(2\eta _{2h}+\eta
_{1h})k_{h}(x_{2\tau _{2}}+x_{4\tau _{2}})x_{1\tau _{2}} \\
\noalign{\medskip}f_{3} & = & -\gamma _{1l}x_{3}-(\eta _{1l}+\eta
_{2l})k_{l}(x_{2}+x_{4})x_{3}-(1-\eta _{1l}-\eta _{2l})\beta
_{l}(x_{1}+x_{3})x_{3}+ \\
\noalign{\medskip} &  & +2e^{-\gamma _{1l}\tau _{3}}(1-\eta _{1l}-\eta
_{2l})\beta _{l}(x_{1\tau _{3}}+x_{3\tau _{3}})x_{3\tau _{3}}+ \\
\noalign{\medskip} &  & +\eta _{1l}e^{-\gamma _{1l}\tau _{3}}k_{l}(x_{2\tau
_{3}}+x_{4\tau _{3}})x_{3\tau _{3}}-b_{1}x_{3}x_{5}l_{1}(x_{3}+x_{4}) \\
\noalign{\medskip}f_{4} & = & -\gamma _{2l}x_{4}+A_{l}(2\eta _{2l}+\eta
_{1l})k_{l}(x_{2\tau _{4}}+x_{4\tau _{4}})x_{3\tau
_{4}}-b_{2}x_{4}x_{5}l_{1}(x_{3}+x_{4}) \\
\noalign{\medskip}f_{5} & = &
a_{1}-a_{2}x_{5}-a_{3}x_{5}l_{2}(x_{4})+2^{n_{1}}a_{4}x_{5\tau
_{5}}l_{2}(x_{4\tau _{5}})%
\end{array}%
\end{equation}%
} We denote $X_{\tau }=X(t-\tau ),$ where $X=(x_{1},x_{2},x_{3},x_{4},x_{5})$. The history is introduced through:
\begin{equation}
X(\theta )=\varphi (\theta ),\ \theta \in \lbrack -\tau _{\max },0],\ \tau
_{\max }=\max (\tau _{1},\tau _{2},\tau _{3},\tau _{4},\tau _{5}).   
\end{equation}%
Two feedback rates are considered in the model: the rate of self-renewal $%
\beta (x)=\beta _{0}\frac{\theta _{1}^{m}}{\theta _{1}^{m}+x^{m}}$ and the
rate of differentiation (through symmetric or asymmetric division) $%
k(x)=k_{0}\frac{\theta _{2}^{n}}{\theta _{2}^{n}+x^{n}}.$ Because
self-renewal is activated by signals from stem cell population and
differentiation is activated by signals from mature cell population,
it is assumed that $\beta $ depends on the sum of stem cell
populations ($x_{1}+x_{3}$), while $k$ depends on the sum of mature cell
population ($x_{2}+x_{4}$). In this way, competition between healthy and CML
cell populations is considered here (see also \cite{jtb}).

In this model, $\gamma _{1}$ and $\gamma _{2}$ are mortality rates, $\beta
_{0}$ and $k_{0}$, the maximal rates of self-renewal, respectively of
asymmetric division or differentiation into leukocyte line, $\theta _{1}$
and $\theta _{2}$ the values for which $\beta $, respectively $k$ attains
half of their maximum value, $A$ an amplification factor of mature cells due
to differentiation, $m$ is the parameter controlling the sensitivity of the
rate $\beta $ to changes in the size of stem populations in $G_{0}$ and $n$
is the parameter controlling the sensitivity of the rate $k$ to changes in the
size of mature population.

The fifth equation of the system models the anti-leukemia T cell immune
response. We assume that after encountering a leukemic cell, a T cell has
two alternatives: either inhibits the leukemic cell and activates a feedback
function to stimulate the production of new T cells, or it is inhibited
itself by the leukemic cell.

The interaction between T cells and CML cells was modeled through the
functions $l_{1}(y)=\frac{1}{b_{3}+y}$ and $l_{2}(y)=\frac{y}{b_{4}+y^{2}}%
. $ Hence, the last terms of the third and fourth equations stand for the
inhibition of leukemic cells by anti-leukemia T cells. The function $%
l_{1}(y) $ was chosen starting from the assumption that the inhibition of
the leukemic cell populations by the immune cells increases with the number
of CML cells up to a certain level and then reaches a maximal value of
inhibition. A further increase in the leukemic population will not modify
this value.

As CML cell population stimulates anti-leukemia immune response only if
leukemia cell population has values in a certain range, called "the optimal
load zone" (see \cite{kim}) and inhibits the immune response if it exceeds a
certain threshold value, to model the influence of CML cell population on T cell
evolution, the feedback function $l_{2}(y)$ was chosen. Consequently, the
third and the fourth terms of the fifth equation represents the rate at
which naive T cells leave and re-enter the effector state after finishing
the minimal developmental program of $n_{1}$ cell divisions, due to antigen
stimulation. These terms reflect the process of production of T cells
influenced by the CML cells.

The following parameters are related to the evolution of T cell population: $%
b_{1}$ represents the loss of leukemic stem cells due to cytotoxic T cells; $%
b_{2}$ is the loss of mature leukemic leukocytes due to cytotoxic T
cells; $b_{3}$ represents a standard half-saturation in a Michaelis-Menten law $%
l_{1}(y)$; $b_{4}$ is a standard half-saturation in a feedback function $%
l_{2}(y)$; $a_{1}$ represents the anti-leukemia T-cell natural supply rate; $%
a_{2}$ represents the anti-leukemia T-cell death rate; $a_{3}$ is
the probability that a T cell survives the encounter with a leukemia cell; $%
a_{4}$ represents the coefficient of influence of the regulatory process; $%
n_{1}$ is the number of antigen depending divisions.

System (\ref{sys}) has four possible types of equilibria with
non-negative components, namely: $E_{0}=(0,0,0,0,x_{5}^{\ast })$, $%
E_{1}=(x_{1}^{\ast },x_{2}^{\ast },0,0,x_{5}^{\ast })$, $E_{2}=(x_{1}^{\ast
\ast },x_{2}^{\ast \ast },x_{3}^{\ast \ast },x_{4}^{\ast \ast },x_{5}^{\ast
\ast })$ and $E_{3}=(0,0,\tilde{x}_{3},\tilde{x}_{4},\tilde{x}_{5})$,
corresponding, from a medical point of view, to an extinction phase, to a
healthy situation, to a chronic and respectively to an aggravated phase.
Note that depending on the parameters values, the system can exhibit one or
more equilibria of type $E_{3}$ ($E_{3.i},i=1,p$ - where the value of p
depends on the biologically relevant range for cell populations).

The positive components of the equilibria $E_{1}$ satisfy:
\[
\gamma _{1h}+[\eta _{1h}(1-e^{-\gamma_{1h}\tau _{1}})+\eta
_{2h}]k_{h}(x_{2}^{\ast })-(1-\eta _{1h}-\eta _{2h})(2e^{-\gamma_{1h}\tau
_{1}}-1)\beta _{h}(x_{1}^{\ast })=0  \label{ep1}
\]
\[
-\gamma _{2h}x_{2}^{\ast }+A_{h}(2\eta _{2h}+\eta _{1h})k_{h}(x_{2}^{\ast
})x_{1}^{\ast }=0 \\
a_{1}-a_{2}x_{5}^{\ast }=0
\]
and the positive components of $E_{3}$ satisfy
\[
\gamma _{1l}+[\eta _{1l}(1-e^{-\gamma_{1l}\tau _{3}})+\eta _{2l}]k_{l}(%
\tilde{x} _{4})-(1-\eta _{1l}-\eta _{2l})(2e^{-\gamma_{1l}\tau _{3}}-1)\beta
_{l}(\tilde{x} _{3})-\gamma_{1}\tilde{x}_{3}\tilde{x_5}l_{1}(\tilde{x}_{3}+%
\tilde{x}_{4})=0 \\
\]
\[
-\gamma _{2l}\tilde{x}_{4}+A_{l}(2\eta _{2l}+\eta _{1l})k_{l}(\tilde{x}_{4})
\tilde{x}_{3}-b_{2}\tilde{x}_{4}\tilde{x}_{5}l_{1}(\tilde{x}_{2}+\tilde{x}
_{4})=0  
\]
\[
a_{1}-a_{2}\tilde{x}_{5}-\tilde{x}_{5}l_{2}(\tilde{x}
_{4})(a_{3}+2^{n_{1}}a_{4})=0
\]
For a detailed theoretical analysis of the stability properties of
the equilibrium points, please see \cite{MESA} and \cite{bad}.

\section{THE FRACTIONAL-ORDER DDE MODEL FOR CML}

In general, three different definitions of fractional derivatives, which are
in general non-equivalent, are widely used: the Gr\"{u}nwald-Letnikov
derivative, the Riemann-Liouville derivative and the Caputo derivative (see
for example \cite{Podlubny}). The main advantage of the Caputo derivative is
that it only requires initial conditions given in terms of integer-order
derivatives, representing well-understood features of physical situations
and thus making it more applicable to real world problems. Highly remarkable
scientific books which provide the main theoretical tools for the
qualitative analysis of fractional-order dynamical systems, and at the same
time, show the interconnection as well as the contrast between classical
differential equations and fractional differential equations, are \cite%
{Podlubny}, \cite{Kilbas}, \cite{Lak}.

In the following, let us give the definition of the Caputo derivative and
introduce the fractional-order DDEs system for CML.

\begin{definition}
For a continuous function $f$, with $f\in L_{loc}^{1}(\mathbb{R}^{+})$, the
Caputo fractional-order derivative of order $q\in (0,1)$ of $f$ is defined
by
\[
^{c}D^{q}f(t)=\frac{1}{\Gamma (1-q)}\int_{0}^{t}(t-s)^{-q}f^{\prime }(s)ds~,
\]
where the gamma function is defined, as usual, as $\Gamma
(z)=\int_{0}^{\infty }e^{-t}t^{z-1}dt~.$
\end{definition}

\begin{remark}
When $q\rightarrow 1$, the fractional order derivative $^{c}D^{q}f(t)$
converges to the integer-order derivative $f^{\prime }(t)$.
\end{remark}

The fractional order DDE model for CML using the Caputo derivative is
{%
\begin{equation}
\begin{array}{lcl}
^{c}D^{q}x_{1} & = & f_{1}(x_{1},x_{2},x_{3},x_{4},x_{1\tau _{1}},x_{2\tau
_{1}},x_{3\tau _{1}},x_{4\tau _{1}}) \\
\noalign{\medskip}^{c}D^{q}x_{2} & = & f_{2}(x_{2},x_{1\tau_{2}},
x_{2\tau _{2}},x_{4\tau _{2}}) \\
\noalign{\medskip}^{c}D^{q}x_{3} & = &
f_{3}(x_{1},x_{2},x_{3},x_{4},x_{5},x_{1\tau _{3}},x_{2\tau _{3}},x_{3\tau
_{3}},x_{4\tau _{3}}) \\
\noalign{\medskip}^{c}D^{q}x_{4} & = & f_{4}(x_{3},x_{4},x_{5},x_{2\tau
_{4}},x_{3\tau _{4}},x_{4\tau _{4}}) \\
\noalign{\medskip}^{c}D^{q}x_{5} & = & f_{5}(x_{4},x_{5},x_{4\tau
_{5}},x_{5\tau _{5}}).%
\end{array}
\label{frac CML}
\end{equation}%
}Although the stationary points of the fractional-order DDEs system (\ref%
{frac CML}) are the same as in the case of the DDEs system (\ref{sys}),
regarding their stability properties, things might be much different (\cite{kas}).
Nevertheless, as for fractional-order differential equations with time-delay
only finite time stability results are known until now (see \cite{laz}, \cite%
{zha}) and the general stability theory is not yet developed, in the
following section, the dynamical behavior of the DDE model and its
fractional counterpart will be analyzed and compared in view of the
numerical results and simulations.

\section{NUMERICAL SIMULATIONS}

In this section, for the healthy cell populations we use the values of the
parameters from \cite{col}, modifying some values of the CML cells
parameters in view of medical evidences attesting that the values of
leukemic parameters are increased or decreased in the case of leukemia.
Accordingly, we consider for leukemic cells a smaller fraction of asymmetric
division, a bigger fraction of self-renewal, lower rates of apoptosis for
stem and mature cells, a bigger rate of self-renewal of stem cells, a bigger
rate of asymmetric division or differentiation of stem cells and a bigger
amplification factor of mature cells. Furthermore, we considered that healthy
cells have a propensity to asymmetric division, while CML cells have a
predilection to self renewal (%
\cite{Tomasetti2010}). Obviously, as these values should be correlated with
the features of the patient and his disease, there are many possible
configurations of parameter space.

In all simulations, we choose the same set of values for the following
parameters: $\beta _{0h}=1.77\cdot days^{-1}$, $k_{0h}=0.1\cdot days^{-1}$ ,
$\theta _{1h}=\theta _{1l}=0.5\cdot 10^{6}cells\ kg^{-1}$, $\theta
_{2h}=\theta _{2l}=36\cdot 10^{8}cells\ kg^{-1}$, $m=4$, $n=3$, $\gamma
_{1h}=0.1\cdot days^{-1}$, $\eta _{1h}=0.7$, $\eta _{1l}=0.1$, $\eta
_{2h}=0.1$, $\eta _{2l}=0.7$, $\gamma _{2h}=2.4$, $A_{h}=829$, $b_{1}=0.3$, $%
b_{2}=0.6$, $b_{3}=36$, $b_{4}=36$, $a_{1}=3$, $a_{2}=0.23$, $a_{3}=0.3$,$%
a_{4}=0.9$, $n_{1}=2,$ $\tau _{1}=2.8$, $\tau _{2}=3.5$, $\tau _{3}=2.1$, $%
\tau _{4}=2.8$, $\tau _{5}=1.4$. For the rest of the parameters, in the following
simulations, we consider two configurations of parameters describing two different
forms of CML: configuration 1, with $\beta _{0l}=2$%
, $k_{0l}=0.4$, $\gamma _{1l}=0.04$, $\gamma _{2l}=1.5,$ $A_{l}=2764.8$ and
configuration 2, corresponding to a more serious disease, with $\beta
_{0l}=2.27$, $k_{0l}=0.8$, $\gamma _{1l}=0.01$, $\gamma _{2l}=0.15,$ $%
A_{l}=5529.6$. The values of the equilibrium points for the two
configurations are listed in the table below.%

\begin{table}
\caption{Equilibrium points for the two configurations of parameters}
\begin{tabular}{|c|c|c|}
\hline
\textbf{Equilibrium} & \textbf{Configuration 1} & \textbf{Configuration 2}
\\ \hline
$E_{0}$ & $(0,0,0,0,13.04)$ & $(0,0,0,0,13.04)$ \\ \hline
$E_{1}$ & $(0.13,4.46,0,0,13.04)$ & $(0.13,4.46,0,0,13.04)$ \\ \hline
$E_{2}$ & $(0.09,0.23,0.06,27.76,25.76)$ & $(0.15,0.25,0.004,32.04,22.98)$
\\ \hline
$E_{3.1}$ & $(0,0,0.06,27.61,25.88)$ & $(0,0,0.92,177.86,14.18)$ \\ \hline
$E_{3.2}$ & $(0,0,0.56,62.46,16.88)$ & $(0,0,0.004,31.98,23.01)$ \\ \hline
\end{tabular}%
\end{table}

Because the aim of this study is to compare the dynamical behavior of the
model's trajectories in the case of the DDEs model and the modified
fractional counterpart, in the following, numerical results and simulations
pertaining to the stability properties of equilibria of type $E_{1},E_{2},$%
and $E_{3}$ will be presented. The history used in examples is constant and
is taken in a neighborhood of the non trivial steady state.

\begin{figure}[htbp]
\centering
\begin{tabular}{cc}
\includegraphics[height=0.42\textheight ]{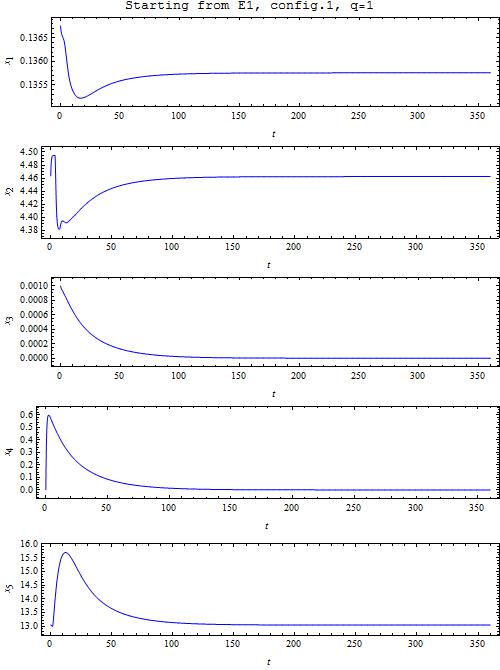} &
\includegraphics[height=0.42\textheight ]{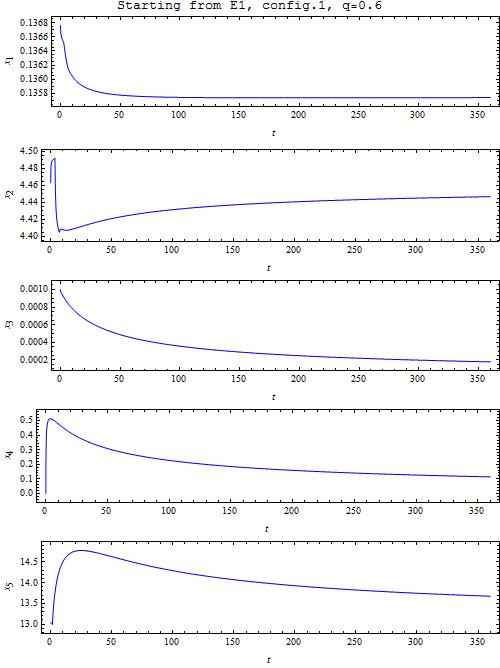}\\
\end{tabular}
\caption{Configuration 1. Dynamics of hematopoietic populations starting from the neighborhood of $E_{1}$ for the integer-order DDEs model (left) and for the fractional-order DDEs model with $q=0.6$ (right).}
\end{figure}

\begin{figure}[htbp]
\centering
\begin{tabular}{cc}
\includegraphics[height=0.42\textheight ]{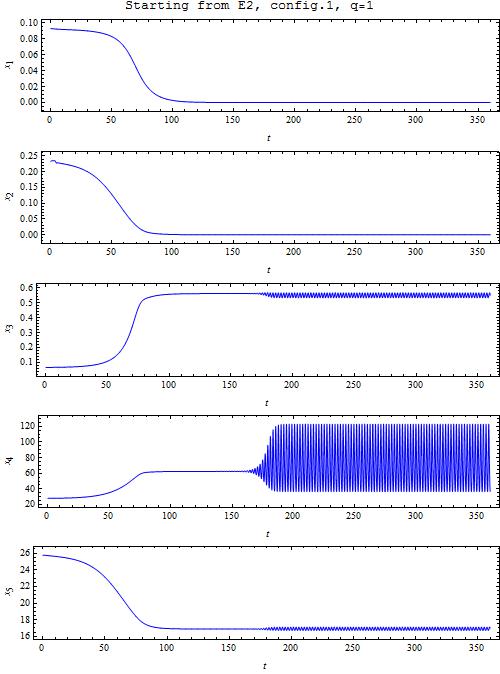} &
\includegraphics[height=0.42\textheight ]{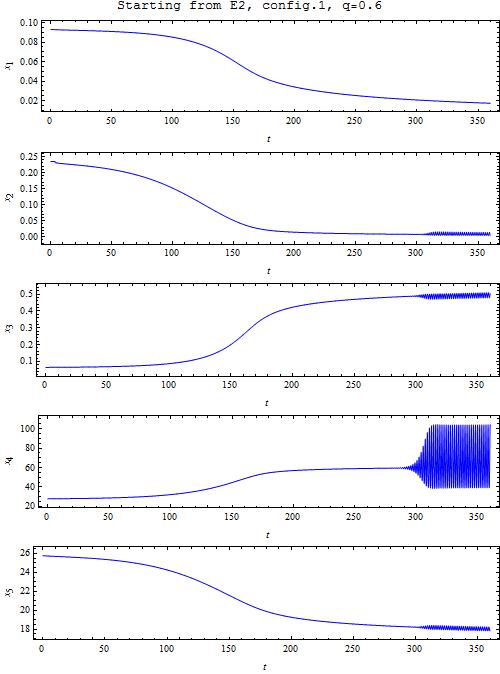}\\
\end{tabular}
\caption{Configuration 1. Dynamics of hematopoietic populations starting from the neighborhood of $E_{2}$ for the integer-order DDEs model (left) and for the fractional-order DDEs model with $q=0.6$ (right).}
\end{figure}

\begin{figure}[htbp]
\centering
\begin{tabular}{cc}
\includegraphics[height=0.42\textheight ]{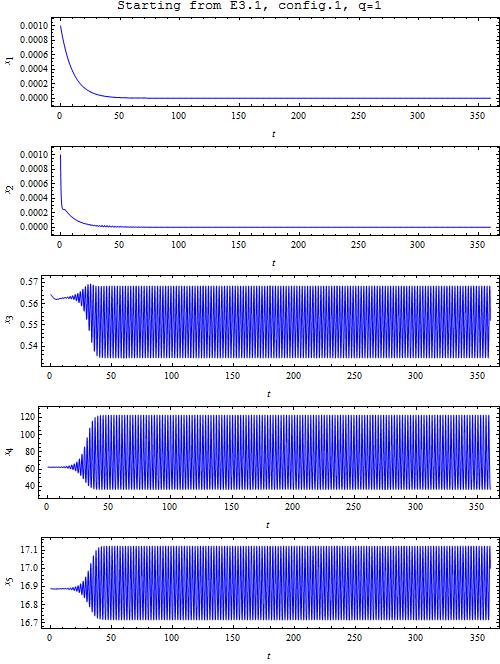} &
\includegraphics[height=0.42\textheight ]{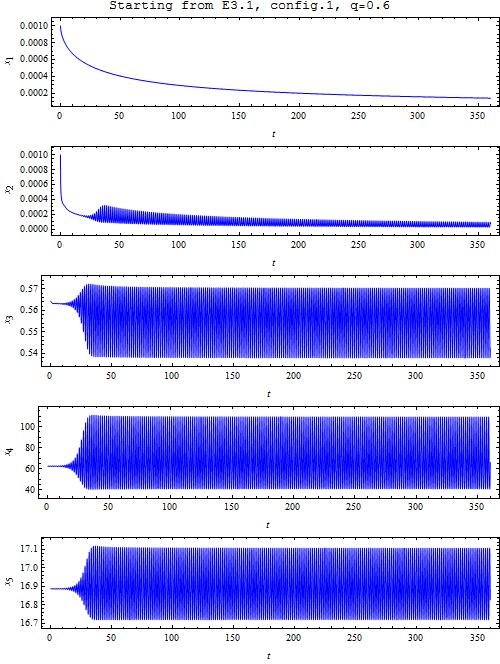}\\
\end{tabular}
\caption{Configuration 1. Dynamics of hematopoietic populations starting from the neighborhood of $E_{3.1}$ for the integer-order DDEs model (left) and for the fractional-order DDEs model with $q=0.6$ (right).}
\end{figure}

\begin{figure}[htbp]
\centering
\begin{tabular}{cc}
\includegraphics[height=0.42\textheight ]{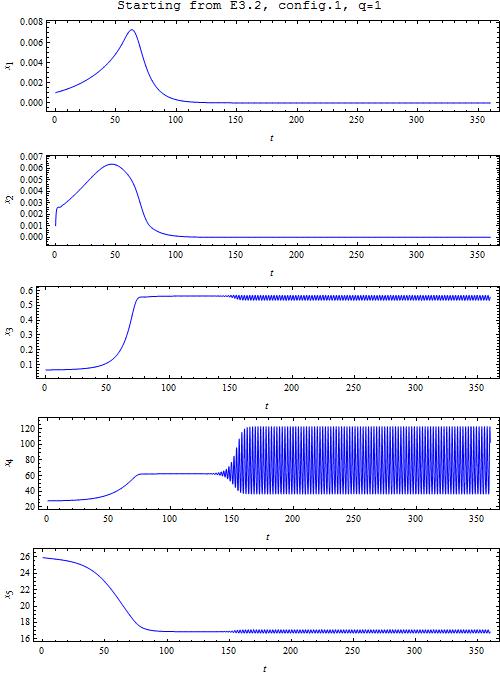} &
\includegraphics[height=0.42\textheight ]{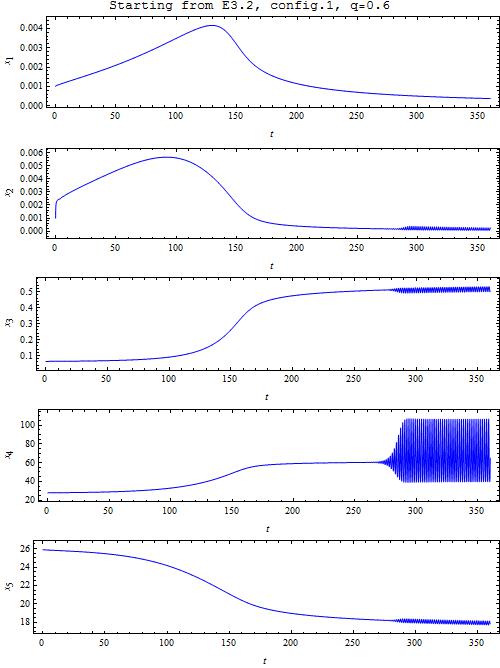}\\
\end{tabular}
\caption{Configuration 1. Dynamics of hematopoietic populations starting from the neighborhood of $E_{3.2}$ for the integer-order DDEs model (left) and for the fractional-order DDEs model with $q=0.6$ (right).}
\end{figure}

\begin{figure}[htbp]
\centering
\begin{tabular}{cc}
\includegraphics[height=0.42\textheight ]{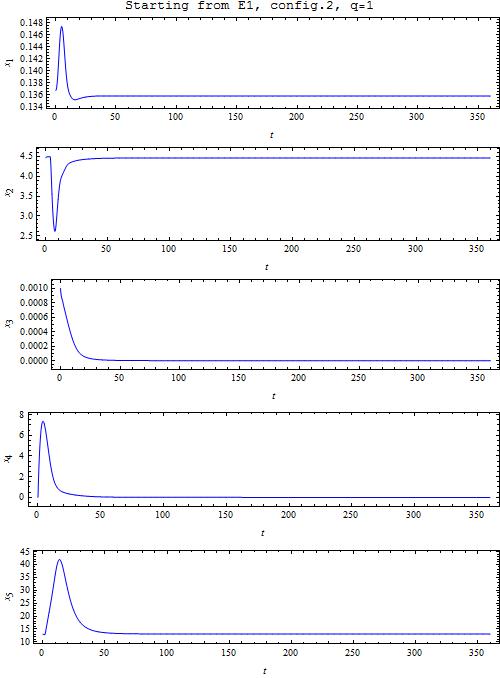} &
\includegraphics[height=0.42\textheight ]{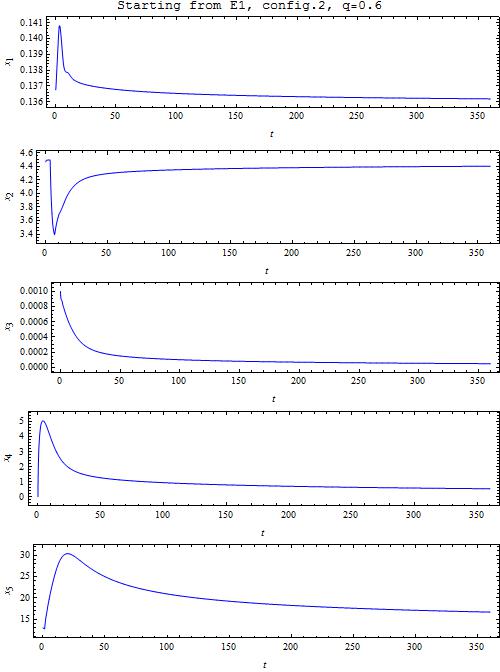}\\
\end{tabular}
\caption{Configuration 2. Dynamics of hematopoietic populations starting from the neighborhood of $E_{1}$ for the integer-order DDEs model (left) and for the fractional-order DDEs model with $q=0.6$ (right).}
\end{figure}

\begin{figure}[htbp]
\centering
\begin{tabular}{cc}
\includegraphics[height=0.42\textheight ]{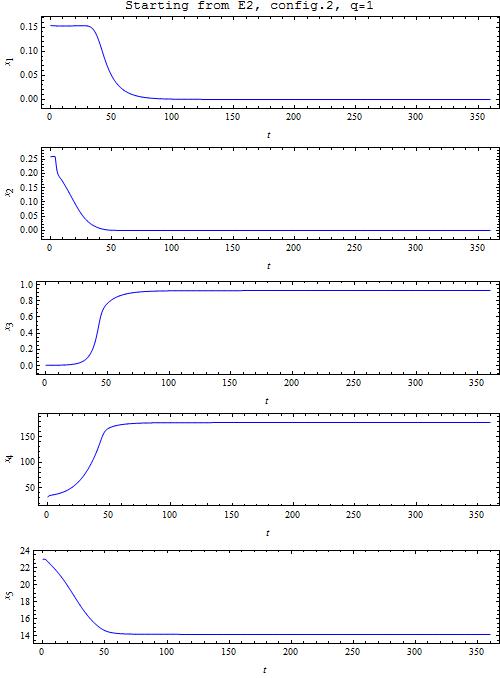} &
\includegraphics[height=0.42\textheight ]{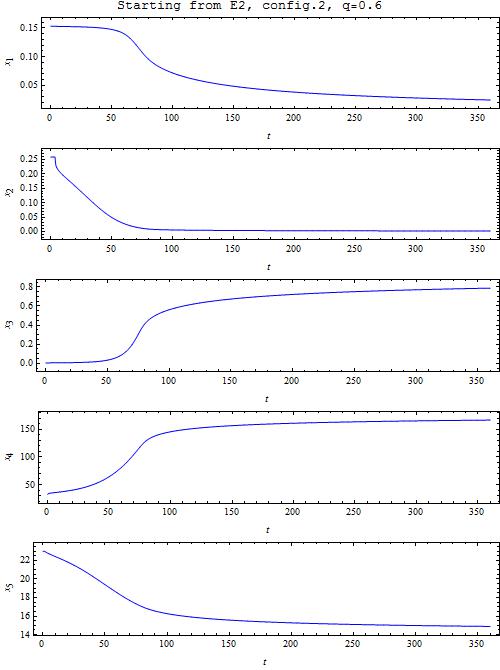}\\
\end{tabular}
\caption{Configuration 2. Dynamics of hematopoietic populations starting from the neighborhood of $E_{2}$ for the integer-order DDEs model (left) and for the fractional-order DDEs model with $q=0.6$ (right).}
\end{figure}

\begin{figure}[htbp]
\centering
\begin{tabular}{cc}
\includegraphics[height=0.42\textheight ]{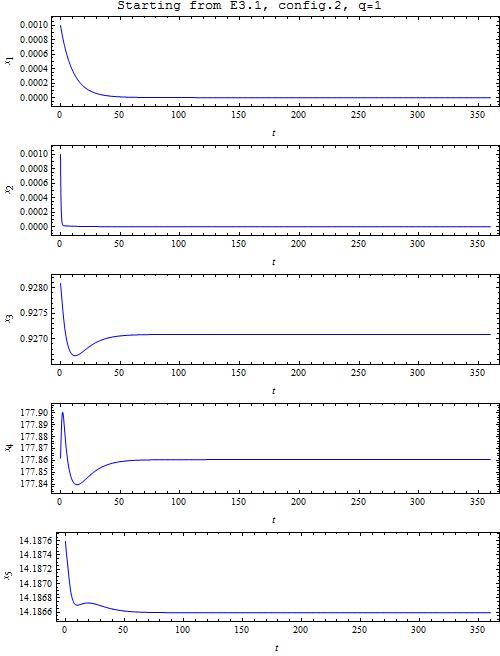} &
\includegraphics[height=0.42\textheight ]{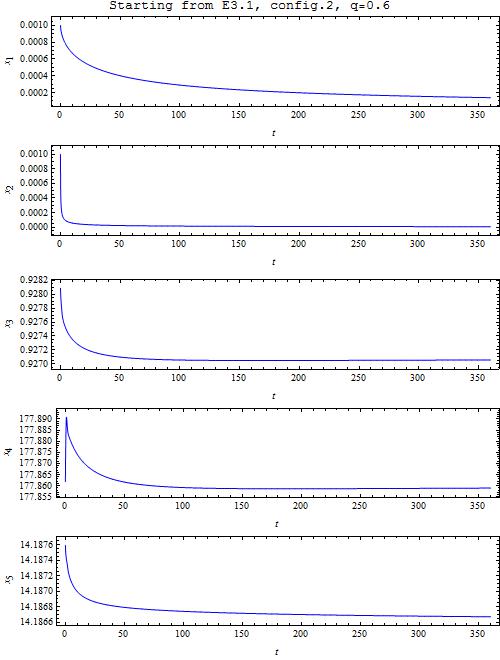}\\
\end{tabular}
\caption{Configuration 2. Dynamics of hematopoietic populations starting from the neighborhood of $E_{3.1}$ for the integer-order DDEs model (left) and for the fractional-order DDEs model with $q=0.6$ (right).}
\end{figure}

\begin{figure}[htbp]
\centering
\begin{tabular}{cc}
\includegraphics[height=0.42\textheight ]{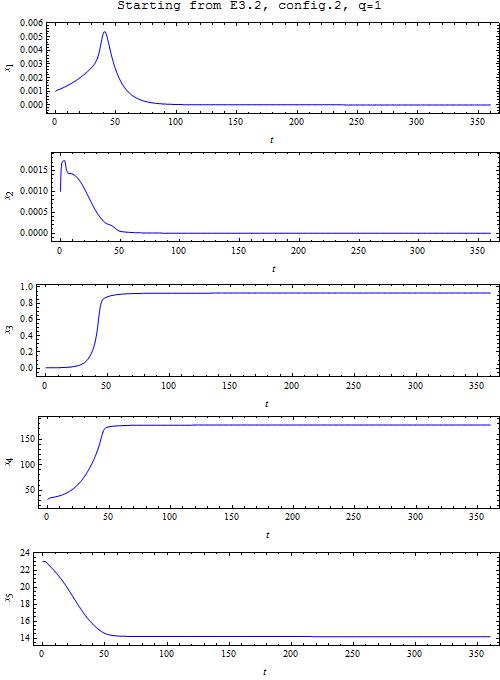} &
\includegraphics[height=0.42\textheight ]{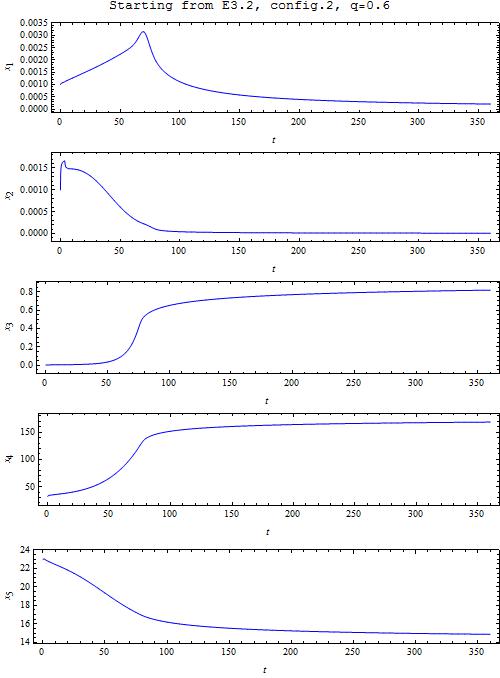}\\
\end{tabular}
\caption{Configuration 2. Dynamics of hematopoietic populations starting from the neighborhood of $E_{3.2}$ for the integer-order DDEs model (left) and for the fractional-order DDEs model with $q=0.6$ (right).}
\end{figure}

\section{CONCLUDING REMARKS}

For both configurations of parameters, we found that equilibrium $E_{1}$ is
stable and as the fractional order $q$ decreases the evolution is slower;
equilibrium $E_{2}$ is unstable, and as the fractional order decreases, one
can notice that the time period needed for the amount of the leukemic populations
to rise up to a certain level become longer ($2-3$ months for
$q=1$ and $5-6$ months for $q=0.6$ in the case of the first configuration and
around $1$ months for $q=1$ and $3$ months for $q=0.6$ for the second configuration).
For the two configurations of parameters values, two steady states
of type $E_{3}$ can coexist: the first one (denoted $E_{3.1}$) can be either
stable or unstable; the second one (denoted $E_{3.2}$) is always unstable.
Equilibrium $E_{3.1}$ is unstable for the first configuration and stable for
the second one, but the time evolution is very similar in the DDEs model and
its fractional counterpart; equilibrium $E_{3.2}$ is unstable in both configurations,
it seems to be attracted by $E_{3.1}$ for the second configuration and as the fractional
order decreases, a very similar behavior as in the case of $E_{2}$ is noticed.

Consequently, an important conclusion can be drawn from the numerical
simulations, as medical evidences regarding CML attest that the disease typically
begins in the chronic phase and progresses to an accelerated phase over the course
of several years, not only in few months. Hence, the results above are important
findings suggesting that the introduction of fractional derivative into the DDEs
system with a small fractional order $q$, might bring a more accurate representation
of the reality into the model and improve it.

\section*{ACKNOWLEDGEMENTS}
This work was supported by grants of the Romanian National Authority for Scientific Research and Innovation, CNCS-UEFISCDI, project no.
PN-II-ID-PCE-2011-3-0198 (D. C\^{a}ndea)  and project no. PN-II-RU-TE-2014-4-0270 (I.R. R\u{a}dulescu and E. Kaslik).

\label{pagefin}

\end{document}